\documentclass[twocolumn,epj]{svjour}
\usepackage{graphics}

\usepackage{graphicx}
\usepackage{epsfig}
\usepackage{ulem}
\usepackage{morefloats}
\usepackage{rotating}
\usepackage{color,xcolor}
\usepackage{soul}
\usepackage{amsmath}
\usepackage{amssymb}
\usepackage{longtable,tabu}
\usepackage{multirow}

\RequirePackage{doi}
\usepackage{hyperref}
\usepackage[sort&compress,numbers]{natbib}

\def\beq{\begin{equation}}
\def\eeq{\end{equation}}
\def\beqa{\begin{eqnarray}}
\def\eeqa{\end{eqnarray}}
\def\ban{\begin{eqnarray*}}
\def\ean{\end{eqnarray*}}
\def\bi{\begin{itemize}}
\def\ei{\end{itemize}}

\def\be {\begin{equation}}
\def\ee {\end{equation}}
\def\bea {\begin{eqnarray}}
\def\eea {\end{eqnarray}}
\def\bc {\begin{center}}
\def\ec {\end{center}}
\def\nn {\nonumber}

\begin{document}
\title{Inner crust equations of state for CompOSE}
\author{Tuhin Malik \and Helena Pais}
%
\mail{Helena Pais, \texttt{hpais@uc.pt}}
\institute{CFisUC, Department of Physics, University of Coimbra,
   3004-516 Coimbra, Portugal}
\date{Received: date / Revised version: date}
%
\abstract{
In this paper, we present some relativistic mean-field inner crust equations of state that have recently been uploaded in the CompOSE online repository. These equations of state fulfill experimental and microscopic constraints, and are also able to reproduce two solar-mass stars. We integrate the TOV equations to obtain the mass-radius relation, and we also calculate the tidal deformability, compactness, and effective tidal deformability to compare with the latest astrophysical data from NICER and LIGO and Virgo. }

%
\maketitle
\section{Introduction}
\label{intro}

Neutron stars (NS) are born in core-collapse supernova (CCSN) events, and, along with black holes (BH), are one of the most compact objects in the Universe. In 2017, the detection of gravitational wave signals from the collision of two NS with the LIGO and Virgo interferometers \cite{Abbott2017},
followed up by the gamma-ray burst GRB170817A and the electromagnetic transient AT2017gfo, set the beginning of this new and very exciting multi-messenger era for the astronomy, nuclear, gravitational and astrophysics community. Later, in 2019, a second signal, GW190425, was detected, a larger system than those of any binary NS known to date \cite{Abbott2020}. Recently, the NICER collaboration has published new radius and mass measurements from PSRJ0030+0451 \cite{nicer} which have been able to set new constrains in neutron star matter. 

The equation of state (EoS) for stellar matter, that should be consistent with theoretical, experimental and observational constraints, is the essential ingredient to build the stars' mass-radius relation. Ideally one would want a unified EoS, so that the whole range of densities, from the outer core to the inner core, could be described. Another approach would be to consider an EoS for each layer of the star, and then match each piece at the transition points. Usually three different layers are considered: outer crust, inner crust and core. In the literature, one can find several outer crust EoS, such as the Baym-Pethick-Sutherland (BPS) EoS \cite{BPS}, the Haensel and Pichon (HP) EoS \cite{HP}, or the R\"uster {\it et al} (RHS) EoS \cite{ruester06}, and it has been shown that the mass-radius relation is independent on the choice of such outer crust EoS \cite{Fortin16}. On the other hand, the NS radius should be sensitive to the inner crust EoS.  In this layer of the star, as the density increases, it is believed that light \cite{Typel2010} and heavy (pasta) \cite{pasta83,pais2012PRL,grill14} nuclei structures are formed, that can modify the neutrino transport, and have an impact on the NS dynamics and cooling \cite{Arcones08}. In principle, light clusters should only appear for temperatures above 1 MeV \cite{Typel2010}, but they are abundant in other high-temperature astrophysical sites, like CCSN or NS mergers. However, commonly used EoS in CCSN simulations, like \cite{LS1991,Shen1998,Shen2010}, only include $\alpha-$clusters even though other light nuclear species should in principle be formed \cite{Raduta2010,Horowitz2006,Hempel2010}.

Microscopic calculations for the inner crust date back to the work of Negele and Vautherin \cite{NV}, where nuclear matter was described within a Hartree-Fock approximation and the geometries were computed using  the Wigner-Seitz (WS) method. This WS method was later used by Douchin and Haensel \cite{DH} for their non-relativistic unified EoS within the  compressible liquid drop model with the SLy4 interaction. In Ref.~\cite{Vinas2021}, they also used the WS approximation in in a variational approach with proton shell and pairing corrections included for the calculation of the inner crust.

Pairing effects in the inner crust have also been discussed, check e.g.\cite{Baldo2005,Haensel2007,Gogelein2007,Chamel2008,Grill2012,Pearson2022} and references therein. Pairing may have some effects on the structure (geometries) in the crust, but overall the effect on the EoS is negligible \cite{Pearson2022}. In Ref.~\cite{Baldo2005}, the authors used different energy functionals and reached the conclusion that the pairing effects do depend on the chosen model, even though they found that paring has an affect on the structure of the inner crust within the WS approximation, see also \cite{Chamel2008}. In Ref.~\cite{Gogelein2007}, pairing correlations were also anaysed in the BCS framework, and they found that the local pairing gaps for neutron pairing 
are significantly smaller in the regions of clusterized matter than in the low-density background gas. 

Recent works on the equation of state (EoS) for sub-saturation matter and the crust-core transition \cite{grill14,Avancini2017,Pais2015,Pais2018} have indicated that it is important to consider unified EoS as this has an effect on the NS radius \cite{Fortin16,Pais2016-vlasov}, and that clusters should be present in the EoS for CCSN and NS mergers, as they may have an effect on the cooling or mass ejection.

To compute these heavy clusters, there are several techniques in the literature, such as quantum \cite{Watanabe2009}, semiclassical \cite{Horowitz2005,Schneider2016} or classical \cite{Dorso2012} Molecular Dynamics calculations, 3D Hartree-Fock calculations \cite{Newton2009,pais2012PRL}, Thomas-Fermi (TF) calculations
\cite{Shen2002,Avancini2010,Maruyama,Okamoto2013,Sharma2015,Bertolino2015}, or extended TF with Skyrme BSk effective potentials \cite{Pearson2018,Perot2019}. In this work, we are going to use the TF approximation to describe the pasta phases inside WS cells with RMF nuclear interactions.

We present eight EoS for the inner crust, where clusterized matter is expected to appear due to the competition between the strong and Coulomb forces. These EoS are all publicly available in the CompOSE database \cite{compose}, a free online repository for EoS that was started during the NewCompStar COST Action, and it is now in the process of being improved and updated. Such EoS can then be used in CCSN and binary NS mergers simulations. 

This paper is organized as follows: a brief summary of the formalism applied to calculate the NS inner crust is given in the next Section, the mass-radius relation from the TOV equations together with the calculation of the tidal deformability is mentioned in Section \ref{sec:mr}, some results are shown in Section \ref{sec:results}, and, finally, in Section \ref{sec:conclusions}, some conclusions are drawn.

\section{The Thomas-Fermi approximation}
\label{sec:1}

We use the Thomas-Fermi approximation \cite{grill14} (and refs therein) to describe the non-homogeneous clusterized matter present in the inner crust of neutron stars. We will use relativistic mean-field models, both with non-linear couplings, like NL3$\omega\rho$L55, TM1e, FSU2, FSU2R, FSU2H, and with density-dependent couplings, like TW, DD2 and DDME2.

We start from the Lagrangian density, 
\begin{equation}
\mathcal{L}=\sum_{i=n,p}\mathcal{L}_{i}+\mathcal{L}_e\mathcal{\,+L}_{{\sigma }}%
\mathcal{+L}_{{\omega }}\mathcal{+L}_{{\rho
}}\mathcal{+L}_{{\gamma }}
\mathcal{+L}_{{nl}} \, , \label{lagdelta}
\end{equation}
where the nucleons are described with vacuum mass $M$ coupled
to the scalar meson $\sigma$ with mass $m_\sigma$,  the vector
isoscalar meson $\omega$ with mass $m_\omega$ and the vector
isovector meson $\rho$ with mass $m_\rho$. We also include a system of electrons with
mass $m_e$, and they interact with the protons through the electromagnetic field $A^{\mu}$. 
The nucleon Lagrangian reads
\begin{equation}
\mathcal{L}_{i}=\bar{\psi}_{i}\left[ \gamma _{\mu }iD^{\mu }-M^{*}\right]
\psi _{i}  \label{lagnucl},
\end{equation}
with
\begin{equation}
iD^{\mu }=i\partial ^{\mu }-g_{\omega}\omega^{\mu }-\frac{g_{\rho }}{2}{\boldsymbol{\tau}}%
\cdot \boldsymbol{\rho}^{\mu } - e \frac{1+\tau_3}{2}A^{\mu} \, ,
\end{equation}
and $M^{*}=M-g_{\sigma}\sigma$ is the Dirac effective mass of the nucleons. $\boldsymbol \tau$ are the SU(2) isospin matrices.
The electron Lagrangian is given by
\begin{equation}
\mathcal{L}_e=\bar \psi_e\left[\gamma_\mu\left(i\partial^{\mu} + e A^{\mu}\right)
-m_e\right]\psi_e.
\label{lage}
\end{equation}
The meson and electromagnetic Lagrangian densities are given by
\begin{eqnarray*}
\mathcal{L}_{{\sigma }} &=&\frac{1}{2}\left( \partial _{\mu }\sigma \partial %
^{\mu }\sigma -m_{\sigma}^{2}\sigma ^{2} -\frac{1}{3}\kappa \sigma ^{3}-\frac{1}{12}%
\lambda \sigma ^{4}  \right)  \\
\mathcal{L}_{{\omega }} &=&-\frac{1}{4} \Omega _{\mu \nu }
\Omega ^{\mu \nu }+ \frac{1}{2}m_{\omega}^{2}\omega_{\mu }\omega^{\mu } + \frac{\xi}{4!} g_{\omega}^{4}(\omega_{\mu}\omega^{\mu })^{2}  \\
\mathcal{L}_{{\rho }} &=&-\frac{1}{4}
\mathbf{R}_{\mu \nu }\cdot \mathbf{R}^{\mu
\nu }+ \frac{1}{2}m_{\rho }^{2}\boldsymbol{\rho}_{\mu }\cdot \boldsymbol{\rho}^{\mu } \\
\mathcal{L}_{{\gamma }} &=&-\frac{1}{4}F _{\mu \nu }F^{\mu  \nu }
\end{eqnarray*}
where $\Omega _{\mu \nu }=\partial _{\mu }\omega_{\nu }-\partial
_{\nu }\omega_{\mu }$, $\mathbf{R}_{\mu \nu }=\partial _{\mu
}\boldsymbol{\rho}_{\nu }-\partial _{\nu }\boldsymbol{\rho}_{\mu
}-\Gamma_{\rho }(\boldsymbol{\rho}_{\mu }\times
\boldsymbol{\rho}_{\nu })$ and $F_{\mu \nu }=\partial _{\mu
}A_{\nu }-\partial _{\nu }A_{\mu }$. 


The non-linear term in the Lagrangian density, $\mathcal{L}_{{nl}}$, that  mixes  the $\omega$, and $\mathbf{\rho}$ mesons, is only present in the models FSU2, FSU2R, FSU2H, NL3$\omega\rho$L55, and TM1e:
\begin{eqnarray}
\mathcal{L}_{{nl}} &=&\Lambda_\omega g_\omega^2 g_\rho^2 \omega_{\nu
}\omega^{\nu } \boldsymbol{\rho}_{\mu }\cdot
\boldsymbol{\rho}^{\mu } \, .
\label{FLTL}
\end{eqnarray}

The models DD2, TW, and DDME2 have instead density-dependent couplings. Their isoscalar couplings of the mesons $i$
to the baryons  are given by  
\begin{eqnarray}
g_{i}(n_B)=g_{i}(n_0)a_i\frac{1+b_i(x+d_i)^2}{1+c_i(x+d_i)^2} \, ,
  i=\sigma, \omega, 
\end{eqnarray}
and the isovector  meson-nucleons coupling by
\begin{eqnarray}
g_{\rho}(n_B)=g_{\rho}(n_0)\exp{[-a_\rho(x-1)]} \, .
\label{grho}
\end{eqnarray}
In the last expressions, $n_0$ is the model-dependent symmetric nuclear saturation density, and $x=n_B/n_0$, with $n_B$ the baryonic density.

The basic principle of the Thomas-Fermi approximation of non-uniform $npe$ matter is that the fields can be considered locally constant at each point because they are assumed to vary slowly \cite{Maruyama,pasta1}.
The calculation starts from the grand canonical potential density,
\begin{eqnarray}
\omega&=&\omega(\{f_{i+}\},\{f_{i-}\} ,\sigma,\omega_0,\rho_{30}) \nonumber \\
&=&{\mathcal E}-T{\mathcal S}-\sum_{i=n,p,e}\mu_in_i
\label{gc}
\end{eqnarray}
where $\{f_{i+}\} (\{f_{i-}\})$, $\mu_i$ and $n_i$ stand for the positive (negative)
energy distribution functions, the chemical potential and baryonic density of particle $i$, respectively, and ${\mathcal S}$ and ${\mathcal E}$ are the total entropy and energy densities \cite{Avancini2010}.

The equations of motion for the meson fields (see e.g., Ref.~\cite{pasta1} for details) follow from the variational conditions 
\begin{equation} 
\frac{\delta\Omega}{\delta\sigma(\bf r)}=\frac{\delta\Omega}{\delta\omega_0(\bf r)}=
\frac{\delta\Omega}{\delta\rho_{30}(\bf r)}=0 \ ,
\label{vc}
\end{equation}
where $\Omega=\int d^3r\,\omega$.
The finite-temperature case is a generalization of the zero-temperature case. The meson fields are expanded in a harmonic oscillator basis of one, two or three dimensions in order to solve the Klein--Gordon equations. The appropriate Green's function is also chosen according to the spatial dimension of interest so that the Poisson equation can be solved. Further details can be found in Refs.~\cite{pasta1,Avancini2010}.

\subsection{RMF models}

We briefly review  the RMF models that will be
considered in the present work, in particular, FSU2 \cite{chen},  FSU2R, and FSU2H \cite{FSU2R,tolos2},
NL3$\omega\rho$L55 \cite{Pais2016-vlasov}, TM1e \cite{shen2020,sumiyoshi19},  TW \cite{TW}, DDME2 \cite{ddme2}, and DD2 \cite{Typel2010}. In Table \ref{tab1}, we show some symmetric nuclear matter
properties calculated at saturation density for these models.

\begin{table}[]
\caption{A few symmetric nuclear matter properties for the models used in this work, calculated at saturation density, $n_0$: the binding energy per particle $B/A$, the incompressibility $K$, the skewness, $Q_0$, the symmetry energy $E_{\rm sym}$, the slope of the symmetry energy $L$, and $K_{\rm sym}$. All quantities are in MeV, except for $n_0$ that is given in fm$^{-3}$.}
\label{tab1}
\setlength{\tabcolsep}{4.pt}
      \renewcommand{\arraystretch}{1.4}
\begin{tabular}{cccccccc}
\hline \hline 
Model              & $n_0$           & $B/A$  & $K$   &$Q_0$ & $E_{\rm sym}$    & $L$     & $K_{\rm sym}$ \\
\hline
NL3$\omega\rho$L55    & 0.148           & -16.24 & 270 & 198        & 31.7 & 55.0  & -8            \\
FSU2R              & 0.151           & -16.28 & 238 & -135       & 30.7 & 47.0  & 56            \\
FSU2H              & 0.151           & -16.28 & 238 & -24.6      & 30.5 & 44.5  & 87            \\
FSU2               & 0.151           & -16.28 & 238 & -149       & 37.6 & 113.0 & 25            \\
TM1e               & 0.145           & -16.30 & 281 & -285       & 31.4 & 40.0  & 3.57          \\
DD2                & 0.149           & -16.02 & 243 & 169        & 31.7 & 55.0  & -93.2         \\
DDME2              & 0.152           & -16.14 & 251 & 479        & 32.3 & 51.0  & -87.2         \\
TW                 & 0.153           & -16.25 & 240 & -540       & 32.8 & 55.0  & -125   \\ \hline  
\end{tabular}
\end{table}

All the models that are being considered in this work are able to produce 2$M_\odot$
stars, and their radii are compatible with the latest astrophysical observations by NICER \cite{nicer}, as it will be discussed next.

FSU2 \cite{chen} has been optimized to reproduce experimental data on several properties of finite nuclei. FSU2R and FSU2H \cite{FSU2R,tolos2} were constructed in order to reproduce some experimental constraints, such as the properties of finite nuclei, and kaon production and
collective flow in HIC, and also to be consistent with chiral effective field theory neutron matter calculations. Both of them have the slope of the symmetry energy below 50 MeV, together with all the other models considered in this work, making them consistent with many nuclear experiments \cite{oertel18,tews17,birkhan17}, whereas FSU2 has a larger slope, $L=113$ MeV. This model, FSU2, however, seems to be compatible with recent analysis \cite{Reed21} of the PREXII experiment \cite{Adhikari21}, where $L$ was estimated to be in the range of $106 \pm$ 37 MeV, though other works give $L$ lower values, $L=85.5\pm 22.2$ MeV \cite{Yue21}, $L=53^{+14}_{-15}$ MeV \cite{Essick21}, and $L=54\pm 8$ MeV \cite{Reinhard2021}.

The NL3$\omega\rho$L55 model was constructed from NL3 \cite{NL3} to model the density dependence of the symmetry, keeping the isoscalar properties fixed, because NL3, like FSU2, has a very large slope of the symmetry energy at saturation. However, NL3 should be adequate to study systems that have a density similar to the ones we find in nuclei, i.e. sub-saturation densities. This model gives  a very good description not only for the properties of stable nuclei but
also for those far from the valley of beta stability.  The nuclear properties fitted are the charge radii, the binding energies, and the available neutron radii of several spherical nuclei. The binding energies and charge radii were taken within an accuracy of
0.1\% and 0.2\%, respectively.

TM1e was built in the same spirit of NL3$\omega\rho$L55 from the TM1 model \cite{Shen1998}, that has a rather large slope parameter $L = 110.8$ MeV, predicting too large radii for
neutron stars, by adding an extra interaction term between the vector and isovector mesons that allows to model the density dependence of the symmetry energy. TM1 model  reproduces the gross
properties of nuclear masses and charge radii.

TW \cite{TW}, DD2 \cite{Typel2010}, and DDME2 \cite{ddme2}, are density-dependent models, i.e. their couplings of the mesons to the nucleons are not fixed, and vary with the density. The parameters of DDME2 were adjusted in order to simultaneously reproduce the properties of nuclear matter and binding energies, charge radii, and differences between neutron and proton radii of spherical nuclei.
The TW model, on the other hand, did not consider any fiting
to experimental data of nuclear radii in the determination of the mode1 parameters, though for nuclear masses, good agreement was found, especially near the valley of stability. The DD2 model follows the parametrization DD \cite{typel2005}, and fits the properties of finite nuclei, like binding energies, spin-orbit splittings, charge
and diffraction radii, surface thicknesses, and the neutron skin
thickness of $^{208}$Pb.
 
 \begin{figure*}
\begin{center}
\resizebox{0.90\textwidth}{!}{\includegraphics{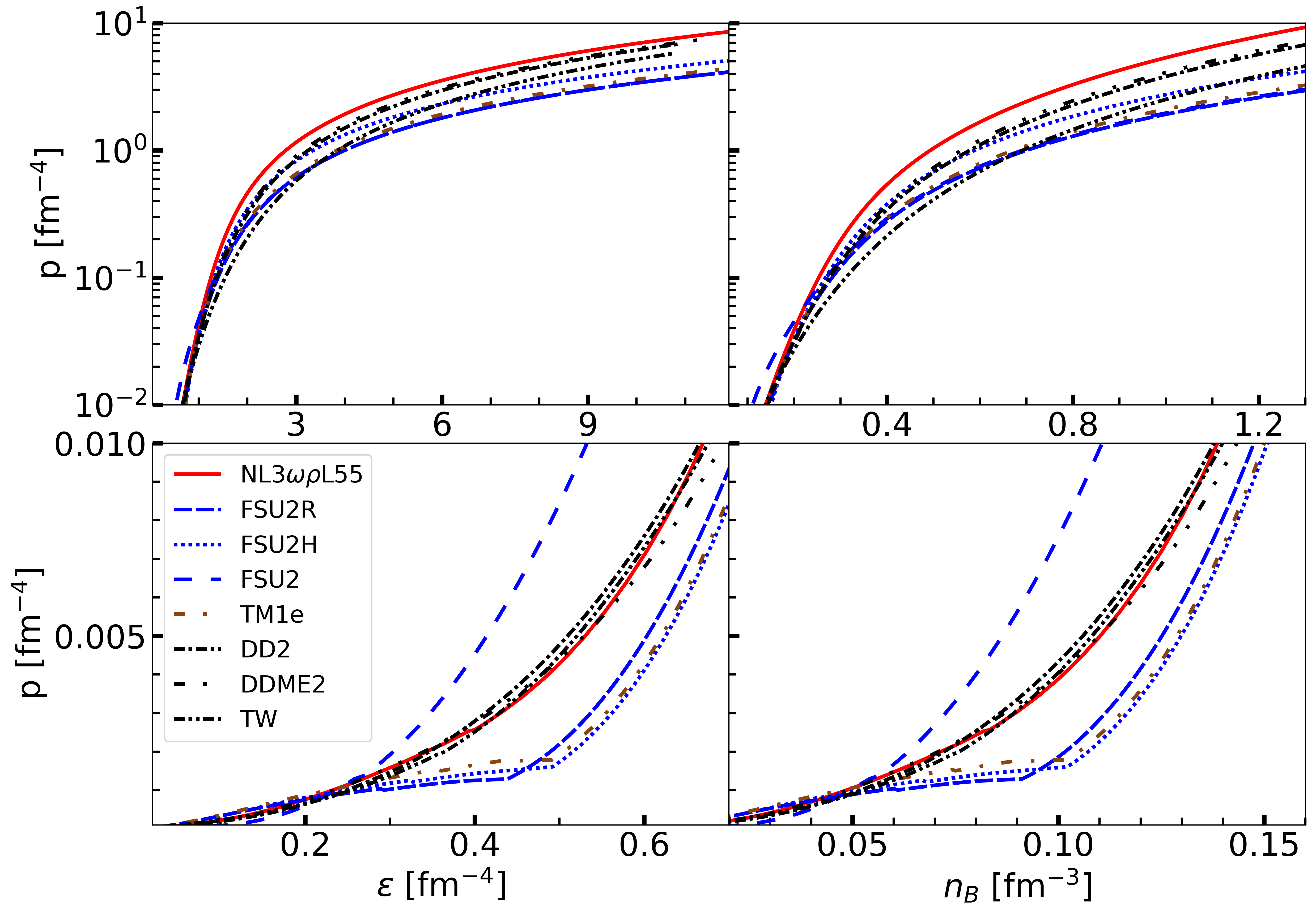}}
\caption{The equation of state, i.e. the pressure as function of the energy density (left) and as a function of the density (right), for the 8 models considered in this work. The bottom panels concentrate in the low-density part of the star, whereas the top panels show the full density range, up to $\sim 8n_0$. }
\label{fig:1}     
\end{center}
\end{figure*}

\section{Neutron Star Properties}
\label{sec:mr}

The NS internal structure depends on the hydrostatic equilibrium between the inward gravitational pull of matter and the outward neutron degeneracy pressure. The first NS model was calculated by Oppenheimer \& Volkoff \cite{Oppenheimer:1939ne} using the exact form of the equations of hydrostatic equilibrium in General Relativity, and which they derived simultaneously with Tolman \cite{Tolman:1939jz} from the Einstein equations.
The TOV equations for the structure of a relativistic spherical and static star are given by
\bea
\label{tov1}
\frac{dP(r)}{dr}=-\frac{G}{r}\frac{\left[\varepsilon+P(r)\right]\left[M(r)+4\pi r^3 P(r)\right ]}{(r-2 GM(r))} \, ,
\eea
\bea
\label{tov2}
\frac{dM(r)}{dr}= 4\pi r^2 \varepsilon \, ,
\eea
with $G$ the gravitational constant. $P(r)$, $\varepsilon(r)$ and $M(r)$ are the pressure, energy density and mass of the neutron star, respectively. These quantities are a function of the distance $r$ from the center of the spherical star. For a given EoS, these equations can be integrated from the origin as an initial value problem for a given choice of central energy density, $(\varepsilon_c)$. The boundary conditions are $P(0) = P_c$ and $M(0) = 0$, with $P_c$ and $M(0)$ the pressure and mass at the center of the NS, respectively. The value of $r=R$, defined as where the pressure vanishes, defines the surface of the star.

The gravitational wave form evolution caused by the tidal deformation of a binary neutron star (BNS) system depends on the EoS of high density matter. The tidal deformability parameter $\lambda$ relates the induced quadrupole moment $Q_{ij}$ of a neutron star due to the strong tidal gravitational field ${\cal E}_{ij}$ of the companion star. This quadrupole deformation in leading order in perturbation is given as  \cite{Flanagan:2007ix,Hinderer:2007mb,Read:2009yp,Read:2013zra,Hinderer:2009ca},
\bea
Q_{ij}=-\lambda{\cal E}_{ij}.
\eea
The parameter $\lambda$ is related to 
the dimensionless tidal Love number  $k_2$ as $k_2=\frac{3}{2} G \lambda R^{-5}$, with $R$ being the radius of the neutron star. 
This parameter $k_2$ can be calculated from the following expression:
\bea
&& k_2 = \frac{8C^5}{5}\left(1-2C\right)^2
\left[2+2C\left(y_R-1\right)-y_R\right]\times \\
&&\bigg\{2C\left(6-3 y_R+3 C(5y_R-8)\right)\nn \\
&&+4C^3\left[13-11y_R+C(3 y_R-2)+2
C^2(1+y_R)\right] \nn \\
&& ~ ~
+3(1-2C)^2\left[2-y_R+2C(y_R-1)\right]\log\left(1-2C\right)\bigg\}^{-1},\nn
\eea
where $C$ $(\equiv m/R)$ is the dimensionless compactness parameter of the star with mass $m$.  The quantity $y_R$ $(\equiv y(R))$ can be obtained by solving
the following differential equation
\bea
r \frac{d y(r)}{dr} + {y(r)}^2 + y(r) F(r) + r^2 Q(r) = 0
\label{TidalLove2} ,
\eea
with
\bea
F(r) = \frac{r-4 \pi r^3 \left( \epsilon(r) - p(r)\right) }{r-2
m(r)},
\eea
\bea
\nonumber Q(r) &=& \frac{4 \pi r \left(5 \epsilon(r) +9 p(r) +
\frac{\epsilon(r) + p(r)}{\partial p(r)/\partial
\epsilon(r)} - \frac{6}{4 \pi r^2}\right)}{r-2m(r)} \\
&-&  4\left[\frac{m(r) + 4 \pi r^3
p(r)}{r^2\left(1-2m(r)/r\right)}\right]^2 \ .
\eea

For a given EoS, Eq.(\ref{TidalLove2}) can be integrated together with the Tolman-Oppenheimer-Volkoff  equations \cite{Weinberg} with the boundary conditions $y(0) = 2$, $p(0)\!=\!p_{c}$ and $m(0)\!=\!0$,{ where} $y(0)$, $p_c$ and $m(0)$ are the dimensionless quantity, pressure and mass 
at the center of the NS, respectively. One can then define the dimensionless tidal deformability, 
$\Lambda = \frac{2}{3}k_2 C^{-5}$. 
The tidal deformabilities of the neutron stars in the BNS system can be combined, and the following weighted average, i.e. the effective tidal deformability, $\tilde \Lambda$, can be calculated  
\bea
\label{lambr}
\tilde \Lambda = \frac{16}{13} \frac{(12 q + 1) \Lambda_1 
+ (12 + q  ) q^4 \Lambda_2}{( 1 + q)^5},
\eea
where $\Lambda_1$ and $\Lambda_2$ are the individual tidal deformabilities corresponding to the two components of the BNS with masses $m_1$ and $m_2$, respectively.  $q=m_2/m_1 < 1$ is the BNS mass ratio.  For $q=1$, the masses are the same, which implies that the individual tidal deformabilities are equal, $\tilde \Lambda = \Lambda_1 = \Lambda_2$. The observable that can be directly measured from GW events is the chirp mass, $\mathcal{M}_c=(m_1m_2)^{3/5}(m_1+m_2)^{-1/5}$.

\section{Results}
\label{sec:results}
 In this Section, we study some of the properties of the RMF models that we have considered in this work. We start by constructing the unified EoS. We then compute the TOV equations to build the mass-radius relations for each model. We then calculate tidal deformabilities and compactness for the systems considered, and confront our findings with recent astrophysical constrains. We also make a comparison with a set of 21 Skyrme EoS, that will be detailed next.

 To construct the unified EoS, we merge the BPS EoS for the outer crust with the TF inner crust EoS for each model, and with the core EoS at the crust-core transition, i.e., the density for which the heavy clusters dissolve. The core EoS consists of homogeneous $npe\mu$ matter in $\beta-$equilibrium at zero temperature. 
These 8 EoS can be found in the CompOSE repository \cite{compose-inner}.  In the Appendix, we also give the inner crust EoS for 3 of the models, namely  TM1e, DD2 and TW, because the other ones have been previously published elsewhere. 

In order to compare our results obtained from unified RMF EoSs, we have also considered 21 diverse Skyrme Hartree–Fock (SHF) EoSs namely the SKa, SKb \cite{Kohler76}, SkI2, SkI3, SkI4 \cite{Reinhard95}, SkI6 \cite{Nazarewicz96}, Sly2, Sly9 \cite{ChabanatPhd}, Sly230a \cite{Chabanat97}, Sly4 \cite{Chabanat98}, SkMP \cite{Bennour89}, SKOp \cite{Reinhard99}, KDE0V1 \cite{Agrawal05}, SK255, SK272 \cite{Agrawal03}, Rs \cite{Friedrich86}, BSk21 \cite{Goriely10}, BSk22, BSk23, BSk24, and BSk25 \cite{Goriely13}. Their nuclear matter properties, such as $K_0$, $Q_0$, $M_0$, $J_0$, $L_0$ and $K_{\rm sym,0}$ vary over a wide range,  as it can be seen from the supplementary material of Ref.~\cite{Alam2016}. The construction performed to obtain these unified Skyrme EoS can be found in Ref.~\cite{Fortin16}.

In Fig.~\ref{fig:1}, we show the pressure as a function of the baryon density and as function of the energy density for all the unified RMF EoSs under consideration in the present work. In the bottom panels, we focus in the low-density range, whereas in the top panels the full range is shown. We see that NL3$\omega\rho$L55 becomes the hardest EoS, when the density becomes above $\sim 0.2$ fm$^{-3}$, whereas below that value, it is FSU2 that shows that behavior, because it has the largest slope of the symmetry energy, $L=113$ MeV, and the lowest incompressibility, $K=238$ MeV. In the low-density regime, the isovector part associated with the symmetry energy is being dominant, making FSU2 the stiffest model, and in the high-density part, it is the isoscalar part associated with the symmetric nuclear matter that prevails and NL3$\omega\rho$L55 becomes the hardest EoS. 
FSU2H, FSU2R and TM1e, having the lowest values of $L$,  show the softest behavior of all the EoS. The density-dependent models are very similar, and show an intermediate behavior as compared to the previous models. Looking to the bottom panels, one can notice a clear change in the EoS slope for the 3 models with the softest behavior, FSU2R, FSU2H and TM1e. These three EoS show an abrupt transition to the core, as compared to the other models. This extreme behaviour  was already observed in Ref.~\cite{Olfa}, and it can also be seen in the neutron matter pressure, where, in the range of densities where the transition to the core happens, even though the incompressibility is always positive, it is in the limit since the pressure presents a plateau.

\begin{figure*}[htp]
\begin{center}
\resizebox{0.95\textwidth}{!}{\includegraphics{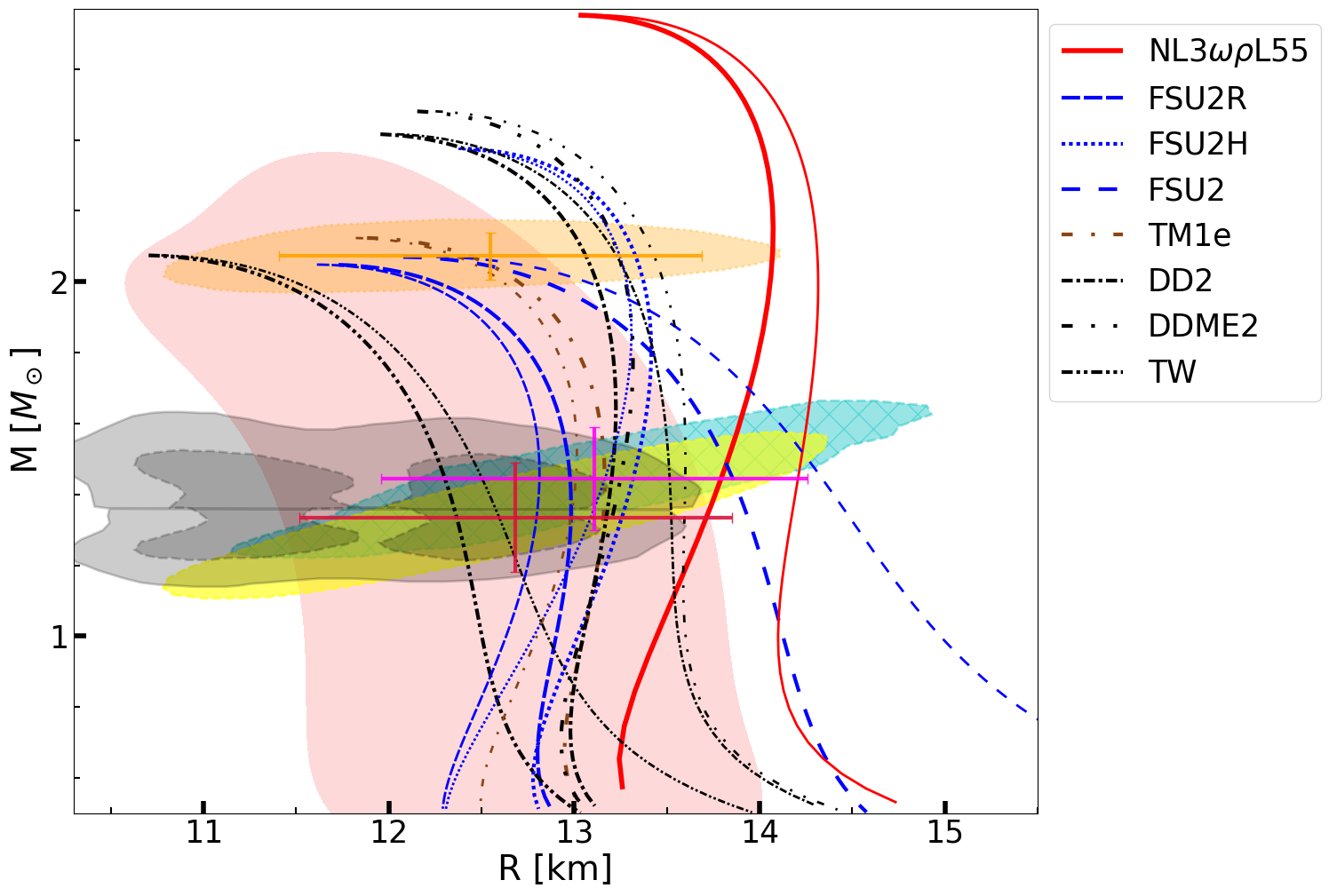}}
\caption{The mass-radius relation for all the RMF unified EoS (thick lines) considered in this work. 
The thin lines correspond to a construction where a BPS-Polytropic fitted crust EoS is used (see text for details). The top (bottom) gray region indicates the heavier
(lighter) NS in the GW170817 event for a parametrized EoS where a lower limit on the maximum mass of $1.97 M_\odot$ was imposed \cite{LIGOScientific:2018cki}, with a 90\% (solid contour) and 50\% (dashed contour) confidence interval. The $1\sigma$ (68\%) confidence region for the 2-D posterior distribution in mass-radii domain from the millisecond pulsar PSR J0030+0451 NICER x-ray data (the cyan hatched and yellow) \cite{nicer} and PSR J0740+6620 \cite{nicer2} (orange) shaded region are also represented. The horizontal (radius) and vertical (mass) error bars represent the $1\sigma$ confidence interval obtained for the 1-D marginalized posterior distribution  of the same NICER data.
 The 21 SHF EoS are represented by the salmon band.}  
\label{fig:2}   
\end{center}
\end{figure*}

In Fig.~\ref{fig:2}, the thick lines represent the mass-radius relation obtained from computing the TOV equations for all the unified RMF EoS considered. As previously stated, we see that all the models attain 2 $M_\odot$ stars. In this Figure, we include the constrains from NICER, given by the $1\sigma$ (68\%) confidence regions, that correspond to the 2-D posterior distribution in mass-radii domain from the  millisecond pulsar PSR J0030+0451 \cite{nicer} (cyan hatched and yellow), and to the   PSR J0740+6620 \cite{nicer2} (orange). The horizontal (radius) and vertical (mass) error bars represent the $1\sigma$ confidence interval obtained for the 1-D marginalized posterior distribution  of the same NICER data. We also represent the constraints from the GW170817 binary event of the heavier (lighter) NS, indicated by the top (bottom) gray region, for a parametrized EoS where a lower limit on the maximum mass of $1.97 M_\odot$ was imposed \cite{LIGOScientific:2018cki}, with a 90\% (solid contour) and 50\% (dashed contour) confidence interval.

One should note that FSU2R, TM1e, DD2 and TW models fall inside the overlap region, given by the darkest color, of the low-mass NS NICER data and the 50\% confidence interval region of the GW event. These models also fit the high-mass NICER constraint. The NL3$\omega\rho$L55, the stiffest EoS, and FSU2, fall outside the 90 \% confidence interval region of the GW170817 constraints, but satisfy the NICER measurements for low-mass NS, with FSU2 satisfying the high-mass constraint as well. In a near future, more precise NICER measurements of joint masses, both low- and high-mass NS, should provide stringent constraints on the EoS.

 In this plot, we also compare our results with EoS built with a different crust treatment, given by the thin lines. The outer crust, however, the Bethe-Pethick-Sutherland (BPS) EoS, is the same as in the unified case. Between the outer crust and the core, we use the polytropic form $p(\varepsilon)=a_1 + a_2 \varepsilon^{\gamma}$ \cite{Carriere:2002bx}, where the parameters $a_1$ and $a_2$ are determined in such a way that the polytropic EoS matches the outer crust at one end ($\rho=10^{-4}$ fm$^{-3}$) and the core at the other end ($\rho=0.04$ fm$^{-3}$). The polytropic index $\gamma$ is taken to be $4/3$. As we can see from the figure, this treatment of the crust EoS causes $\pm4$ \% uncertainty on the radius near the 1.4$M_\odot$ NS, clearly showing that a realistic calculation of the crust makes a difference in the radius, especially for intermediate- to low-mass NS. Another feature is the fact that both the model with the highest incompressibility, NL3$\omega\rho$L55, and the model with the highest slope of the symmetry energy, FSU2, present the highest difference for the radius of a 1.4 $M_\odot$ NS, between the two constructions. This is can also be seen more clearly in Table~\ref{tab2}, that we will discuss next.

For completeness,  in this same figure, we also consider a set of 21 non-relativistic SHF EoS, represented by the salmon band. These EoS have, in general, smaller radii, as compared to the RMF ones, and are also able to reach 2$M_\odot$. However, one should note that the full band of 21 SHF EoSs satisfy the overlap regions of GW and NICER constraints for the low-mass NS, though, on the other hand, half of the band does not fullfill the NICER constraints for the high-mass NS. This is because the SHF EoS have a tendency to give lower radius compared to RMF.

\begin{table*}[htp]
\centering
\caption{ A few NS properties for the models used in this work: $n_t$ the core-crust transition density, $M_{\rm max}$ the NS maximum mass, and $c_s^2$: the central square of the speed of sound for the maximum mass NS. The $R_{1.4}$, $R_{1.4}^{\rm core}$, $R_{1.4}^{\rm crust}$ and $\Lambda_{1.4}$ are the total radius, the radius for core, the crust thickness and the tidal deformability for a 1.4 $M_\odot$ NS respectively. The $R_{1.4}^{\rm BPS+Poly}$ is the radius for a 1.4 $M_{\odot}$ NS calculated with a BPS+Poly crust (see text for details).\label{tab2} } 
\setlength{\tabcolsep}{10.5pt}
      \renewcommand{\arraystretch}{1.4}
\begin{tabular}{cccccccccc}
\hline 
\hline 
\multirow{2}{*}{No} & \multirow{2}{*}{Model} & $n_t$ & M$_{\rm max}$           & $R_{1.4}$ & $R_{1.4}^{\rm BPS+Poly}$ & $R_{1.4}^{\rm core}$ & $R_{1.4}^{\rm crust}$ & $\Lambda_{1.4}$ & $c_s^2$ (at M$_{\rm max}$) \\
                    &                        & {[}fm$^{-3}${]} & {[}M$_\odot${]}  & {[}km{]}  & [km] & {[}km{]}             & {[}km{]}         &                 & {[}$c^2${]} \\ \hline
1  & NL3$\omega\rho$L55 & 0.082 & 2.752 & 13.76 & 14.19 & 12.39 & 1.37   & 930 & 0.799 \\
2  & FSU2R              & 0.083 & 2.048 & 12.98 & 12.80 & 11.82 & 1.16   & 627 & 0.399 \\
3  & FSU2H              & 0.087 & 2.375 & 13.30 & 13.14 & 12.04 & 1.26   & 773 & 0.497 \\
4  & FSU2               & 0.054 & 2.070 & 13.89 & 14.40 & 12.72 & 1.17   & 852 & 0.399 \\
5  & TM1e               & 0.089 & 2.123 & 13.16 & 13.00 & 11.87 & 1.29   & 656 & 0.428 \\
6  & DD2                & 0.067 & 2.416 & 13.19 & 13.52 & 12.03 & 1.16   & 682 & 0.743 \\
7  & DDME2              & 0.072 & 2.480 & 13.23 & 13.60 & 11.99 & 1.24   & 703 & 0.754 \\
8  & TW                 & 0.074 & 2.074 & 12.33 & 12.54 & 11.26 & 1.07   & 402 & 0.697 \\ \hline
\end{tabular}
\end{table*}

In Table \ref{tab2}, we show a few NS properties obtained from our eight sets of unified RMF EoSs, namely the core-crust transition density, $n_t$, the NS maximum mass  $M_{\rm max}$, and the central square of the speed of sound for the maximum mass NS, $c_s^2$. The $R_{1.4}$, $R_{1.4}^{\rm core}$, $R_{1.4}^{\rm crust}$ and $\Lambda_{1.4}$ are the total radius, the radius for core, the crust thickness and the tidal deformability for a 1.4 $M_\odot$ NS respectively. The $R_{1.4}^{\rm BPS+Poly}$ in the table is the radius for a 1.4 $M_{\odot}$ NS calculated with BPS+Poly fitted crust as discussed above. For our RMF EoSs, the crust thickness is, on average $\sim 1.3$ km. These two different crust treatments, the unified and the BPS+Poly, give a difference of $\sim \pm 4\%$ for the radius of 1.4$M_\odot$ NS, this difference increasing with lower-mass NS. As already seen in Fig.~\ref{fig:2}, the models NL3$\omega\rho$L55 and FSU2 present the largest difference, due to their high incompressibility and slope of the symmetry energy. However, this will have little effect on the dimensionless tidal deformability, since this quantity is insensitive to the crust because the Love number $k_2$ compensates the changes of the radii in different crust constructions. The dimensionless tidal deformability of 1.4$M_\odot$ NS obtained for NL3$\omega\rho$L55 and FSU2 is greater than 800, being disfavored by the GW170817 constraint. This will be discussed in detail in the next figures. All these models, being relativistic, are causal, as one can see from the square of the speed of sound for the center of NS maximum mass, also given in the Table.

\begin{figure}
\begin{center}
\resizebox{0.48\textwidth}{!}{\includegraphics{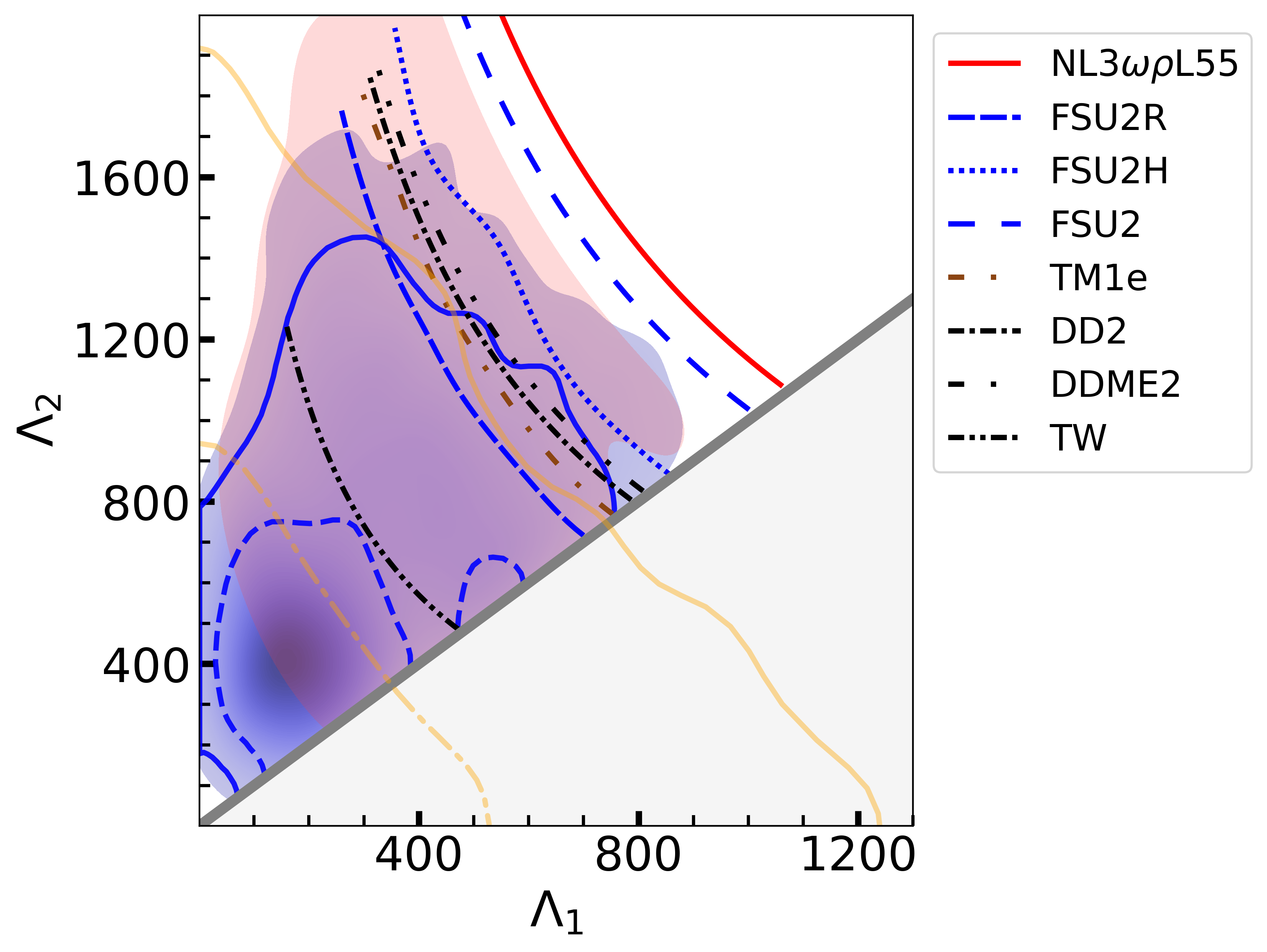}}
\caption{The dimensionless tidal deformability parameters $\Lambda_1$ and $\Lambda_2$ of the binary neutron star merger from the GW170817 event, considering all the RMF EoSs, and using the observed chirp mass of $M_{\rm chirp}=1.186$ M$_\odot$. The orange solid (dashed) line represents the 90\% (50\%) confidence interval, whereas the blue region shows the marginalized posterior using a parametrized EoS with a maximum mass requirement of 1.97 $M_\odot$, with the solid (dashed) lines representing the 90\% (50\%) confidence interval. The salmon band represents the set of the 21 considered SHF EoSs.}
\label{fig:3}   
\end{center}
\end{figure}

In Fig.~\ref{fig:3}, we show the dimensionless tidal deformability parameters $\Lambda_1$ and $\Lambda_2$ for the 2 objects involved in the BNS event from GW170817, with masses $m_1$ and $m_2$. The curves correspond to the EoS considered in this work, and were obtained by varying $m_1$ in the range $1.365 < m_1 < 1.6$ M$_\odot$, and $m_2$ was calculated by keeping the chirp mass fixed at $M_{\rm chirp}=1.186$ M$_\odot$, as observed in the GW170817 event. The orange solid (dashed) line represents the 90\% (50\%) confidence interval from a marginalized posterior for the tidal deformabilities of the two binary components of GW170817 \cite{LIGOScientific:2018cki}. The blue region refers to the marginalized posterior using a parametrized EoS with a maximum mass requirement of at least 1.97 $M_\odot$ \cite{LIGOScientific:2018cki}, with the solid (dashed) lines representing the 90\% (50\%) confidence interval. One can notice that the 90\% confidence interval obtained from GW170817 by LIGO-Virgo with or without the NS maximum mass constraint, very much disfavors the models NL3$\omega\rho$L55 and FSU2. As previously observed in Figs. 1 and 2, these two models are the ones with the hardest EoS behavior, and GW170817 seems to disfavor very stiff EoS \cite{Malik2018}, irrespective of its origin: for NL3$\omega\rho$L55, the stiffness is mostly caused  by the symmetric nuclear matter part (high $K$), whereas for FSU2, it is the symmetry energy (high $L$).  TW and FSU2R, on the other hand, are the only 2 models that fulfill the 90\% confidence level with and without the maximum mass constraint. All the other models, except FSU2H, only fulfill the constraint without the maximum mass.  
 In the same figure, we also plot the  $\Lambda_1$ -  $\Lambda_2$ regions obtained for the 21 SHF sets and for the same chirp mass, $M_{\rm chirp}=1.186$ M$_\odot$. Contrary to Fig.~\ref{fig:2}, where part of the GW and NICER constraints were not fulfilled with these models, when we consider the tidal deformabilities, these EoS do cover most of the regions that overlap with the GW constraints, except in the $\Lambda\lesssim 400$ part of the spectrum.

\begin{figure*}
\begin{center}
\resizebox{0.85\textwidth}{!}{\includegraphics{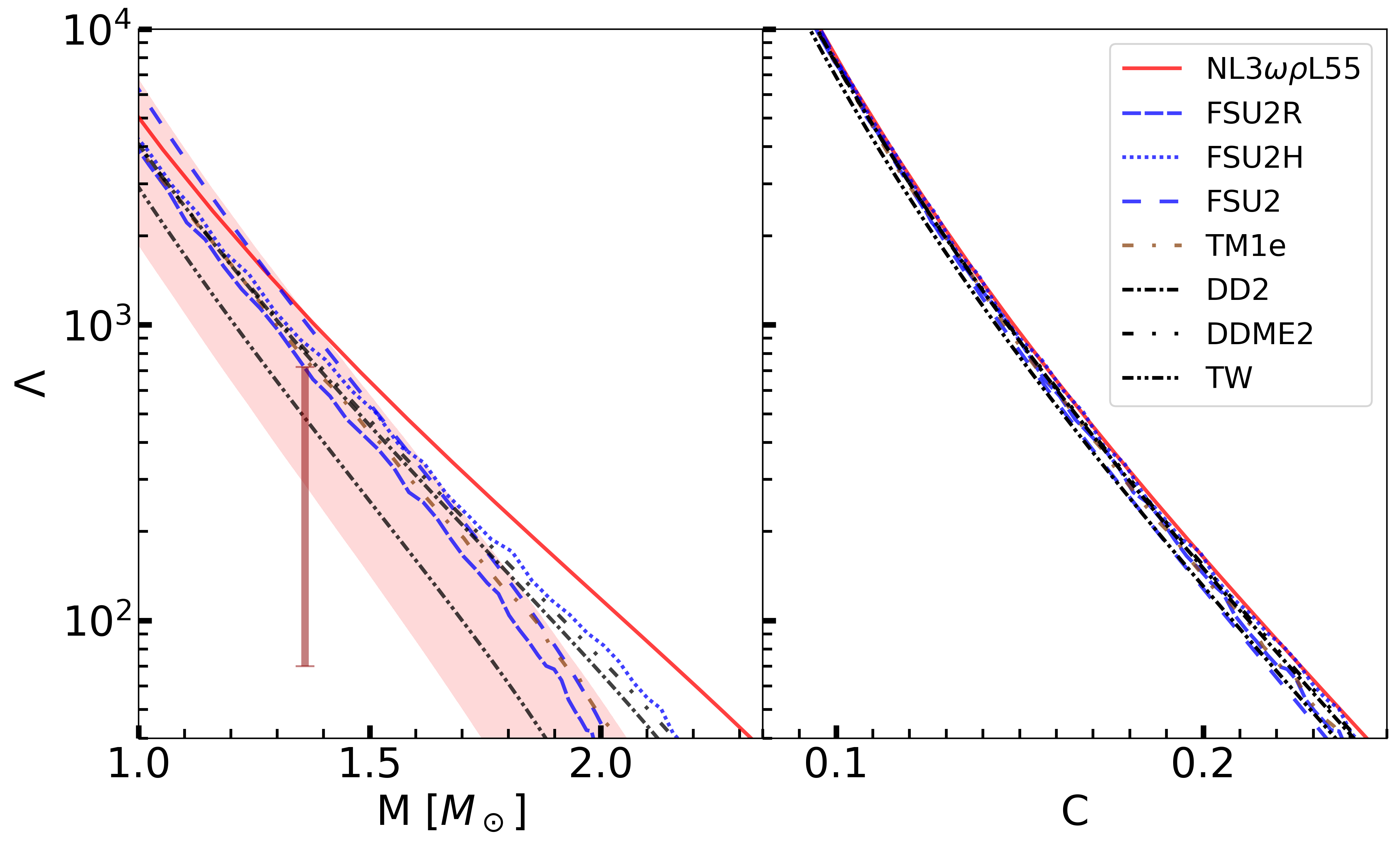}}
\caption{The dimensionless tidal deformability $\Lambda$ as a function of the NS mass $M$ (left), and as a function of the dimensionless compactness parameter $C$ (right) for all the RMF models considered in this work. The vertical line in the left panel represents the 90\% confidence interval obtained for the tidal deformability of a 1.36$M_\odot$ star \cite{LVC2019}. Also in that panel, the salmon band  represents the region covered by the set of 21 SHF EoSs.}
\label{fig:4}    
\end{center}
\end{figure*}

Next we present the variation of the dimensionless tidal deformability with respect to the NS mass and compactness for our set of EoSs. There exists few universal, and nearly universal, relations among star properties. These relations would allow an independent determination of the neutron star radius and tidal deformability. In Fig. \ref{fig:4}, we obtain the dimensionless tidal deformability as a function of the NS mass (left panel), and as a function of dimensionless compactness (right panel) for the set of RMF EoS we have considered.  
The vertical dark red bar in the left panel represents the 90\% confidence interval obtained for the tidal deformability of a $1.36 M_{\odot}$ star in the range $\Lambda_{1.36} \in [70,720]$ \cite{LVC2019}.

The FSU2 and NL3$\omega\rho$L55 models show a deviation from the above constraint due to their large stiffness, as compared to the others. Among the other six RMF EoSs, TW completely satisfies the tidal deformability bounds, whereas the other models have a marginal overlap. In that same panel, the salmon band represents the set of SHF EoS we are considering. A major portion of this band overlaps with the tidal deformability constraint. In the right panel, we extend our investigation to the well-known analytic universal relation "I-Love-C" \cite{Jiang:2020uvb}. This relation is also numerically explored with many different sets of EoS in Ref. \cite{Carson:2019rjx}. In our Figure, we can see that, even though the set of EoS we use is quite small, since we are restricted to 8 EoS, the $\Lambda-C$ relation seems also EoS-insensitive, model-independent.  In the future, the precise and simultaneous determination of NS mass and tidal deformability will help to determine NS radius with more precision in a model-independent way, using this "I-Love-C" relation.

\begin{figure*}
\begin{center}
\resizebox{0.85\textwidth}{!}{\includegraphics{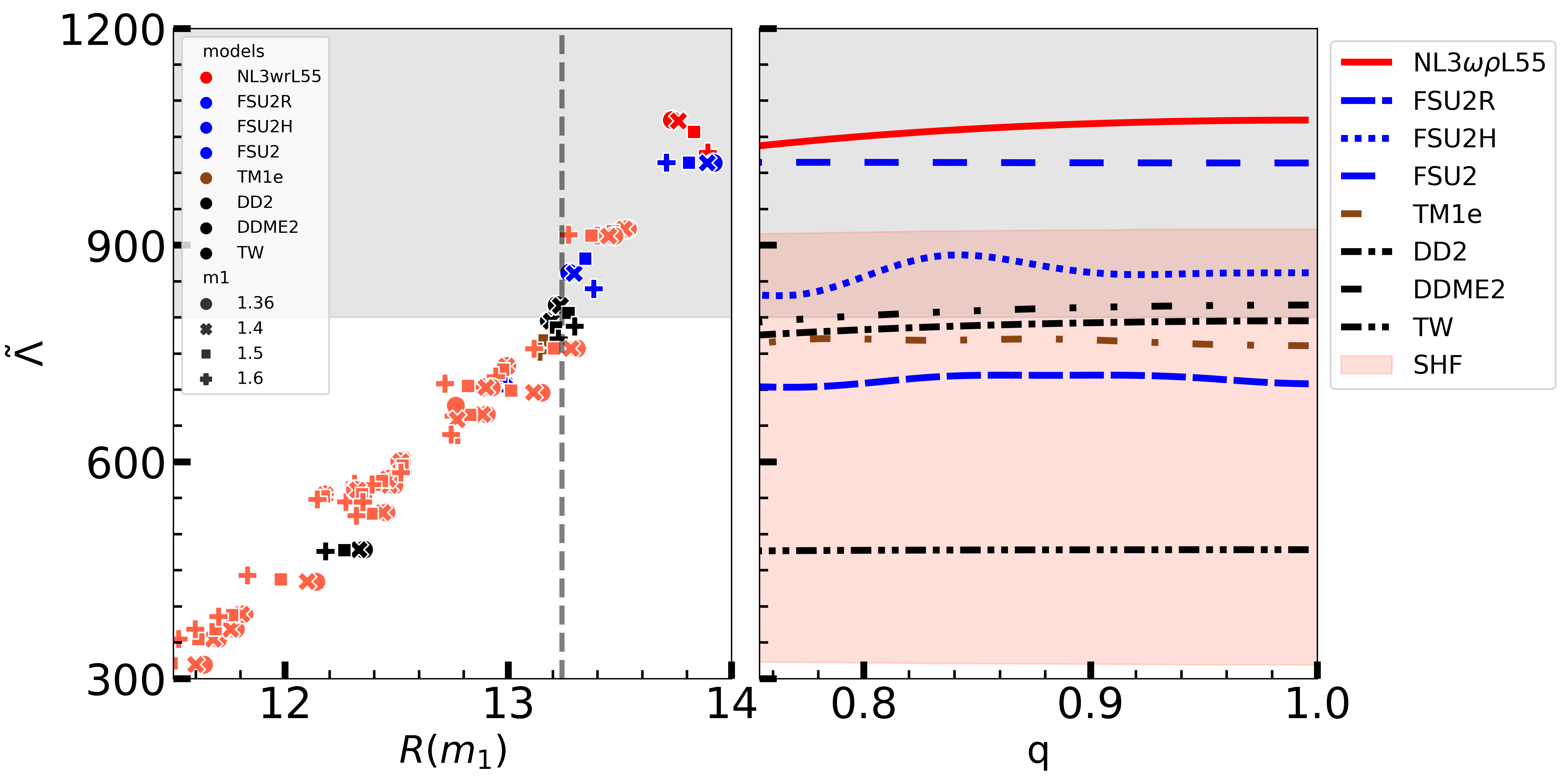}}
\caption{The effective tidal deformability $\tilde \Lambda$ of the binary system from GW170817
as a function of the radius $R$ (left) for different primary neutron star masses $m_1$, and as a function of the mass ratio $q$ (right) for all the RMF EoSs considered in this work. The gray shaded region refers to the excluded region of $\tilde \Lambda$ by the GW170817 event. The vertical dashed line in the left panel is for $R=13.24$ km, the upper bound of the NS radius for the same range of NS $m_1$ masses considered (see text for more details). The salmon colored symbols in the left panel correspond to the SHF EoS, as well as the salmon band in the right panel.}
\label{fig:5}     
\end{center}
\end{figure*}

In Fig.~\ref{fig:5}, we show the effective tidal deformability $\tilde \Lambda$ as a function of the radius $R$ (left) for different primary masses $m1$, that range from 1.36 to 1.6 $M_\odot$, and as a function of the mass ratio $q=m_2/m_1$ of the binary system for all the EoS considered. The chirp mass $M_{\rm chirp}$ is fixed to 1.186 $M_\odot$. The gray shaded band shows the excluded region of $\tilde \Lambda$ by the GW170817 event, i.e. $\tilde \Lambda > 800$ where the symmetric 90\% credible interval was considered \cite{LIGOScientific:2017vwq}.
In the left panel, we can see that the $\tilde \Lambda-R(m_1)$ relation is relatively insensitive to the primary mass component $m_1$, a conclusion also shown before by other authors using different sets of EoS \cite{Raithel:2018ncd}. Moreover, we see that NL3$\omega\rho$L55 and FSU2 are excluded by this constraint, as it has been previously shown in other Figures, as well as FSU2H. The DD2 and DDME2 models lie just at the limit of the allowed region by GW170817, and the other models fit this constraint. We also calculated the radius that correspond to the NS masses ranging from 1.36 to 1.6 $M_\odot$ for each EoS, and we observe that $R$ should  be $ \leq 13.24$ km. 
In the right panel, we see that $\tilde \Lambda$  shows an almost independent behaviour on the mass ratio $q$ for each EoS. FSU2H and FSU2R show some fluctuations though. This may be due to the transition between the crust and the core. As other authors also found \cite{Raithel:2018ncd}, we see from this Figure that the effective tidal deformability weakly depends on the masses, but depends quite a lot on the radii.
The 21 SHF EoS are also represented in this Figure. As seen before, the fact that they attain lower radii than the RMF EoS, makes them attain a lower effective tidal deformability. One can also say that a almost-linear fit could be made in the $\tilde \Lambda-R(m_1)$ relation if both the SHF and RMF EoS were to be taken together.

\section{Conclusions} \label{sec:conclusions}

In this paper, we present a few unified RMF equations of state for neutron star matter, with an inner crust calculated from a TF method where heavy clusters are included. These EoS are given in the free online repository CompOSE. Moreover, they satisfy constraints coming from earth experiments, microscopic neutron matter calculations from $\chi-$EFT models, and they all reproduce 2 $M_\odot$ constraints. We were also interested in checking the effect of these unified EoS in the macroscopic properties of the stars, i.e. the mass and radius, and to inspect if these EoS fit recent astrophysical constraints from NICER and from the GW170817 event, the first gravitational wave signal from the collision of two neutron stars that was measured by LIGO and Virgo in 2017. In this work, we were also interested in understanding this crust construction. For that effect, we built another 8 EoS, using the same RMF models for the core and the BPS EoS for the outer crust, but with a different inner crust construction. A polytropic functional was used to mimic this NS layer. We observed that the difference in the 1.4 $M_\odot$ radii was of $\sim 4\%$, being higher for lower-mass NS. Models with a high slope of the symmetry energy and incompressibility presented the biggest difference in the radius. We also compared our unified RMF EoS with a set of 21 SHF models. These models tend to have lower radii than the RMF models, which make them to attain lower tidal deformabilities.  

Concerning the unified RMF models, a first conclusion we observed is that two of the models under study, NL3$\omega\rho$L55 and FSU2, do not fit the constraint from GW170817 due to the high slope of the symmetry energy (FSU2) or due to the high incompressibility (NL3$\omega\rho$L55). This means that the tidal deformability parameter excludes EoS that are very stiff. Another conclusion was the fact that all the models fit the NICER constraint for the radius of the PSR J0030+0451 pulsar. We were also able to prove the "I-Love-C" relations, even though we are using a small set of EoS. In fact, the dimensionless NS compactness and the tidal deformability seem very well correlated and model-independent. From the calculation of the effective tidal deformability, we were able to prove that this quantity depends very weakly on the NS masses but there's a very high dependence on the NS radius. This points to the fact that we need more (and more accurate) NS radii measurements, something that we hope that the future will soon bring.

\section*{Appendix}
{\onecolumn
\setlength{\tabulinesep}{2.4pt}
\begin{longtable}{c|ccc|ccc}
\caption{The inner crust equation of state for the TM1e, DD2 and TW models.}\tabularnewline
\hline 
\multirow{2}{*}{$n_B$} & \multicolumn{3}{c}{$\cal E$}    & \multicolumn{3}{c}{$\cal P$}         \\ \cline{2-7}  
                       & TM1e       & DD2        & TW         & TM1e       & DD2        & TW         \\
{[}fm$^{-3}${]}        & \multicolumn{3}{c}{{[}fm$^{-4}${]}} & \multicolumn{3}{c}{{[}fm$^{-4}${]}}  \\ \hline  
\endhead     
\hline
\endfoot
\endlastfoot
                       &            &            &            &            &            &            \\
2.0000E-03             & 9.5176E-03 & 9.5307E-03 & -          & 1.1504E-05 & 9.9328E-06 & -          \\
3.0000E-03             & 1.4284E-02 & 1.4302E-02 & 1.4282E-02 & 1.9916E-05 & 1.5811E-05 & 1.6926E-05 \\
4.0000E-03             & 1.9053E-02 & 1.9076E-02 & 1.9049E-02 & 2.9950E-05 & 2.2349E-05 & 2.4781E-05 \\
5.0000E-03             & 2.3825E-02 & 2.3851E-02 & 2.3819E-02 & 4.1454E-05 & 2.9444E-05 & 3.3396E-05 \\
6.0000E-03             & 2.8600E-02 & 2.8628E-02 & 2.8590E-02 & 5.4275E-05 & 3.6893E-05 & 4.2721E-05 \\
7.0000E-03             & 3.3377E-02 & 3.3406E-02 & 3.3363E-02 & 6.8516E-05 & 4.4697E-05 & 5.2603E-05 \\
8.0000E-03             & 3.8156E-02 & 3.8186E-02 & 3.8137E-02 & 8.3719E-05 & 5.2806E-05 & 6.2840E-05 \\
9.0000E-03             & 4.2936E-02 & 4.2966E-02 & 4.2913E-02 & 1.0009E-04 & 6.1320E-05 & 7.3634E-05 \\
1.0000E-02             & 4.7719E-02 & 4.7747E-02 & 4.7689E-02 & 1.1747E-04 & 6.9985E-05 & 8.4834E-05 \\
1.1000E-02             & 5.2504E-02 & 5.2529E-02 & 5.2467E-02 & 1.3576E-04 & 7.9107E-05 & 9.6439E-05 \\
1.2000E-02             & 5.7290E-02 & 5.7312E-02 & 5.7246E-02 & 1.5507E-04 & 8.8533E-05 & 1.0835E-04 \\
1.3000E-02             & 6.2078E-02 & 6.2096E-02 & 6.2026E-02 & 1.7494E-04 & 9.8213E-05 & 1.2066E-04 \\
1.4000E-02             & 6.6867E-02 & 6.6881E-02 & 6.6807E-02 & 1.9577E-04 & 1.0835E-04 & 1.3318E-04 \\
1.5000E-02             & 7.1658E-02 & 7.1666E-02 & 7.1589E-02 & 2.1725E-04 & 1.1889E-04 & 1.4615E-04 \\
1.6000E-02             & 7.6451E-02 & 7.6452E-02 & 7.6372E-02 & 2.3925E-04 & 1.2978E-04 & 1.5948E-04 \\
1.7000E-02             & 8.1244E-02 & 8.1239E-02 & 8.1156E-02 & 2.6195E-04 & 1.4124E-04 & 1.7311E-04 \\
1.8000E-02             & 8.6040E-02 & 8.6026E-02 & 8.5940E-02 & 2.8521E-04 & 1.5305E-04 & 1.8700E-04 \\
1.9000E-02             & 9.0836E-02 & 9.0814E-02 & 9.0726E-02 & 3.0878E-04 & 1.6546E-04 & 2.0134E-04 \\
2.0000E-02             & 9.5634E-02 & 9.5603E-02 & 9.5512E-02 & 3.3280E-04 & 1.7843E-04 & 2.1604E-04 \\
2.1000E-02             & 1.0043E-01 & 1.0039E-01 & 1.0030E-01 & 3.5743E-04 & 1.9186E-04 & 2.3109E-04 \\
2.2000E-02             & 1.0523E-01 & 1.0518E-01 & 1.0509E-01 & 3.8201E-04 & 2.0605E-04 & 2.4639E-04 \\
2.3000E-02             & 1.1003E-01 & 1.0997E-01 & 1.0987E-01 & 4.0689E-04 & 2.2080E-04 & 2.6220E-04 \\
2.4000E-02             & 1.1484E-01 & 1.1476E-01 & 1.1466E-01 & 4.3218E-04 & 2.3616E-04 & 2.7842E-04 \\
2.5000E-02             & 1.1964E-01 & 1.1956E-01 & 1.1945E-01 & 4.5736E-04 & 2.5227E-04 & 2.9494E-04 \\
2.6000E-02             & 1.2444E-01 & 1.2435E-01 & 1.2424E-01 & 4.8270E-04 & 2.6894E-04 & 3.1197E-04 \\
2.7000E-02             & 1.2925E-01 & 1.2914E-01 & 1.2903E-01 & 5.0809E-04 & 2.8653E-04 & 3.2950E-04 \\
2.8000E-02             & 1.3406E-01 & 1.3394E-01 & 1.3382E-01 & 5.3353E-04 & 3.0487E-04 & 3.4734E-04 \\
2.9000E-02             & 1.3886E-01 & 1.3873E-01 & 1.3862E-01 & 5.5917E-04 & 3.2383E-04 & 3.6579E-04 \\
3.0000E-02             & 1.4367E-01 & 1.4353E-01 & 1.4341E-01 & 5.8461E-04 & 3.4374E-04 & 3.8474E-04 \\
3.1000E-02             & 1.4848E-01 & 1.4832E-01 & 1.4820E-01 & 6.1021E-04 & 3.6447E-04 & 4.0425E-04 \\
3.2000E-02             & 1.5329E-01 & 1.5312E-01 & 1.5300E-01 & 6.3570E-04 & 3.8581E-04 & 4.2407E-04 \\
3.3000E-02             & 1.5810E-01 & 1.5792E-01 & 1.5779E-01 & 6.6124E-04 & 4.0821E-04 & 4.4464E-04 \\
3.4000E-02             & 1.6291E-01 & 1.6271E-01 & 1.6259E-01 & 6.8663E-04 & 4.3152E-04 & 4.6552E-04 \\
3.5000E-02             & 1.6772E-01 & 1.6751E-01 & 1.6738E-01 & 7.1181E-04 & 4.5539E-04 & 4.8726E-04 \\
3.6000E-02             & 1.7254E-01 & 1.7231E-01 & 1.7218E-01 & 7.3695E-04 & 4.8042E-04 & 5.0956E-04 \\
3.7000E-02             & 1.7735E-01 & 1.7711E-01 & 1.7698E-01 & 7.6193E-04 & 5.0637E-04 & 5.3247E-04 \\
3.8000E-02             & 1.8217E-01 & 1.8191E-01 & 1.8178E-01 & 7.8682E-04 & 5.3292E-04 & 5.5583E-04 \\
3.9000E-02             & 1.8698E-01 & 1.8672E-01 & 1.8657E-01 & 8.1160E-04 & 5.6064E-04 & 5.8000E-04 \\
4.0000E-02             & 1.9180E-01 & 1.9152E-01 & 1.9137E-01 & 8.3623E-04 & 5.8928E-04 & 6.0488E-04 \\
4.1000E-02             & 1.9661E-01 & 1.9632E-01 & 1.9617E-01 & 8.6050E-04 & 6.1857E-04 & 6.3022E-04 \\
4.2000E-02             & 2.0143E-01 & 2.0112E-01 & 2.0097E-01 & 8.8462E-04 & 6.4902E-04 & 6.5647E-04 \\
4.3000E-02             & 2.0625E-01 & 2.0593E-01 & 2.0577E-01 & 9.0864E-04 & 6.8044E-04 & 6.8343E-04 \\
4.4000E-02             & 2.1106E-01 & 2.1073E-01 & 2.1058E-01 & 9.2233E-04 & 7.1247E-04 & 7.1090E-04 \\
4.5000E-02             & 2.1588E-01 & 2.1554E-01 & 2.1538E-01 & 9.4579E-04 & 7.4572E-04 & 7.3938E-04 \\
4.6000E-02             & 2.2070E-01 & 2.2035E-01 & 2.2018E-01 & 9.6910E-04 & 7.7992E-04 & 7.6857E-04 \\
4.7000E-02             & 2.2552E-01 & 2.2515E-01 & 2.2498E-01 & 9.9216E-04 & 8.1499E-04 & 7.9852E-04 \\
4.8000E-02             & 2.3034E-01 & 2.2996E-01 & 2.2979E-01 & 1.0150E-03 & 8.5072E-04 & 8.2903E-04 \\
4.9000E-02             & 2.3516E-01 & 2.3477E-01 & 2.3459E-01 & 1.0376E-03 & 8.8771E-04 & 8.6060E-04 \\
5.0000E-02             & 2.3998E-01 & 2.3958E-01 & 2.3940E-01 & 1.0601E-03 & 9.2552E-04 & 8.9293E-04 \\
5.1000E-02             & 2.4480E-01 & 2.4439E-01 & 2.4421E-01 & 1.0824E-03 & 9.6429E-04 & 9.2613E-04 \\
5.2000E-02             & 2.4963E-01 & 2.4920E-01 & 2.4901E-01 & 1.1044E-03 & 1.0039E-03 & 9.6008E-04 \\
5.3000E-02             & 2.5445E-01 & 2.5401E-01 & 2.5382E-01 & 1.1262E-03 & 1.0443E-03 & 9.9464E-04 \\
5.4000E-02             & 2.5927E-01 & 2.5883E-01 & 2.5863E-01 & 1.1479E-03 & 1.0857E-03 & 1.0304E-03 \\
5.5000E-02             & 2.6409E-01 & 2.6364E-01 & 2.6344E-01 & 1.1693E-03 & 1.1279E-03 & 1.0669E-03 \\
5.6000E-02             & 2.6892E-01 & 2.6846E-01 & 2.6825E-01 & 1.1905E-03 & 1.1711E-03 & 1.1042E-03 \\
5.7000E-02             & 2.7374E-01 & 2.7327E-01 & 2.7306E-01 & 1.2115E-03 & 1.2151E-03 & 1.1425E-03 \\
5.8000E-02             & 2.7856E-01 & 2.7809E-01 & 2.7787E-01 & 1.2323E-03 & 1.2601E-03 & 1.1817E-03 \\
5.9000E-02             & 2.8339E-01 & 2.8290E-01 & 2.8268E-01 & 1.2529E-03 & 1.3058E-03 & 1.2216E-03 \\
6.0000E-02             & 2.8821E-01 & 2.8772E-01 & 2.8749E-01 & 1.2733E-03 & 1.3492E-03 & 1.2625E-03 \\
6.1000E-02             & 2.9304E-01 & 2.9254E-01 & 2.9230E-01 & 1.2935E-03 & 1.3967E-03 & 1.3042E-03 \\
6.2000E-02             & 2.9786E-01 & 2.9736E-01 & 2.9712E-01 & 1.3136E-03 & 1.4447E-03 & 1.3436E-03 \\
6.3000E-02             & 3.0269E-01 & 3.0218E-01 & 3.0193E-01 & 1.3333E-03 & 1.4941E-03 & 1.3874E-03 \\
6.4000E-02             & 3.0752E-01 & 3.0700E-01 & 3.0675E-01 & 1.3530E-03 & 1.5436E-03 & 1.4320E-03 \\
6.5000E-02             & 3.1234E-01 & 3.1182E-01 & 3.1156E-01 & 1.3724E-03 & 1.5838E-03 & 1.4775E-03 \\
6.6000E-02             & 3.1717E-01 & 3.1664E-01 & 3.1638E-01 & 1.3917E-03 & 1.6360E-03 & 1.5238E-03 \\
6.7000E-02             & 3.2200E-01 & 3.2147E-01 & 3.2120E-01 & 1.4108E-03 & 1.6891E-03 & 1.5712E-03 \\
6.8000E-02             & 3.2682E-01 & -          & 3.2601E-01 & 1.4297E-03 & -          & 1.6193E-03 \\
6.9000E-02             & 3.3165E-01 & -          & 3.3083E-01 & 1.4484E-03 & -          & 1.6683E-03 \\
7.0000E-02             & 3.3648E-01 & -          & 3.3565E-01 & 1.4666E-03 & -          & 1.7064E-03 \\
7.1000E-02             & 3.4131E-01 & -          & 3.4047E-01 & 1.4849E-03 & -          & 1.7590E-03 \\
7.2000E-02             & 3.4614E-01 & -          & 3.4529E-01 & 1.5031E-03 & -          & 1.8118E-03 \\
7.3000E-02             & 3.5097E-01 & -          & 3.5011E-01 & 1.5211E-03 & -          & 1.8653E-03 \\
7.4000E-02             & 3.5580E-01 & -          & 3.5494E-01 & 1.5388E-03 & -          & 1.9194E-03 \\
7.5000E-02             & 3.6062E-01 & -          & -          & 1.5043E-03 & -          & -          \\
7.6000E-02             & 3.6545E-01 & -          & -          & 1.5234E-03 & -          & -          \\
7.7000E-02             & 3.7028E-01 & -          & -          & 1.5424E-03 & -          & -          \\
7.8000E-02             & 3.7511E-01 & -          & -          & 1.5612E-03 & -          & -          \\
7.9000E-02             & 3.7994E-01 & -          & -          & 1.5799E-03 & -          & -          \\
8.0000E-02             & 3.8477E-01 & -          & -          & 1.5983E-03 & -          & -          \\
8.1000E-02             & 3.8960E-01 & -          & -          & 1.6165E-03 & -          & -          \\
8.2000E-02             & 3.9443E-01 & -          & -          & 1.6345E-03 & -          & -          \\
8.3000E-02             & 3.9926E-01 & -          & -          & 1.6523E-03 & -          & -          \\
8.4000E-02             & 4.0410E-01 & -          & -          & 1.6698E-03 & -          & -          \\
8.5000E-02             & 4.0893E-01 & -          & -          & 1.6869E-03 & -          & -          \\
8.6000E-02             & 4.1376E-01 & -          & -          & 1.7038E-03 & -          & -          \\
8.7000E-02             & 4.1859E-01 & -          & -          & 1.7202E-03 & -          & -          \\
8.8000E-02             & 4.2342E-01 & -          & -          & 1.7361E-03 & -          & -          \\
8.9000E-02             & 4.2825E-01 & -          & -          & 1.7514E-03 & -          & -          \\ 
\hline
\end{longtable}}
\twocolumn

\begin{acknowledgement}

We thank C. Provid\^encia for carefully reading the manuscript. This work was partly supported by the FCT (Portugal) Projects No. UID/FIS/04564/2020 and POCI-01-0145-FEDER-029912, and by PHAROS COST Action CA16214. H.P. acknowledges the grant CEECIND/03092/2017 (FCT, Portugal). 

\end{acknowledgement}

\end{document}